\documentclass[fleqn,twoside]{article}
\usepackage{espcrc2}

\usepackage{graphicx}
\usepackage[figuresright]{rotating}

\newcommand{\AmS}{{\protect\the\textfont2
  A\kern-.1667em\lower.5ex\hbox{M}\kern-.125emS}}

\title{Microarcsecond Astrometry using the SKA}

\author{E. Fomalont\address[MCSD]{National Radio Astronomy Observatory \\
        520 Edgemont Road, Charlottesville, VA  22903  USA}
        and M. Reid\address[MCSD]{Center for Astrophysics \\
        60 Garden Street, Cambridge, MA  02138  USA}}

\begin{document}

\begin{abstract}
The sensitivity and versatility of SKA will provide microarcsec
astrometric precision and high quality milliarcsec-resolution images
by simultaneously detecting calibrator sources near the target source.
To reach these goals, we suggest that the long-baseline component of
SKA contains at least 25\% of the total collecting area in a region
between 1000 to 5000 km from the core SKA.  We also suggest a minimum
of 60 elements in the long-baseline component of SKA to provide the
necessary (u-v) coverage.  For simultaneous all-sky observations,
which provide absolute astrometric and geodetic parameters, we suggest
using ten independent subarrays each composed of at least six
long-baseline elements correlated with the core SKA.  We discuss many
anticipated SKA long-baseline astrometric experiments: determination
of distance, proper motion and orbital motion of thousands of stellar
objects; planetary motion detections; mass determination of degenerate
stars using their kinetics; calibration of the universal distance
scale from 10 to $10^7$ pc; the core and inner-jet interactions of
AGN.  With an increase by a factor of 10 in absolute astrometric
accuracy using simultaneous all sky observations, the fundamental
quasar reference frame can be defined to $<10~\mu$as and tied to the
solar-system dynamic frame to this accuracy.  Parameters associated
with the earth rotation and orientation, nutation, and geophysical
parameters, can be accurately monitored.  Tests of fundamental physics
include: solar and Jovian deflection experiments, the sky frame
accuracy needed to interpret the gravity wave/pulsar-timing
experiment, accurate monitoring of spacecraft orbits which impacts
solar system dynamics.

\vspace{1pc}
\end{abstract}

\maketitle

\section{INTRODUCTION}
The position and motion of celestial objects provide a crucial
understanding of many astrophysical problems.  With the impressive
sensitivity of SKA, which will detect extremely faint radio sources,
its design must not preclude the ability to make microarcsec accurate
astrometric measurements of these faint objects.  With this goal in
mind, we discuss the present SKA technical specifications in \S 2, and
suggest modifications which will strengthen its astrometric and
imaging capabilities, as well as its milliarcsec imaging potential.
In \S 3 we describe the most exciting science associated with motion
of celestial objects, and in \S4 we outline some of the astrometric
and geodetic experiments, and conclude with tests of fundamental
physics.  A short summary is given in \S5.

\section{TECHNICAL IMPROVEMENTS FOR THE ASTROMETRIC USE OF THE SKA}

    The present specifications of SKA \cite{jon04} are anchored by
a sensitivity of $A/T=20000$~m$^2$K$^{-1}$.  For many continuum VLBI
observations at 8 GHz will provide good resolution, so
sensitivity calculations will be made at this frequency.  With a
bandwidth of 2 GHz for two polarizations, the RMS noise for SKA in 10
min is about $0.1~\mu$Jy and in 12 hours about $0.013~\mu$Jy.  For
comparison, the VLA sensitivity is $A/T=220$~m$^2$K$^{-1}$ at 8 GHz
(The EVLA will be 370) with a maximum bandwidth of 100 MHz in each of
two polarizations.  Hence, the sensitivity of SKA for wide-band
continuum observations at 8 GHz will be more than 400 times that of
the VLA\footnote{see
http://www.vla.nrao.edu/astro/guides/exposure/calc.html using 200 MHz
effective bandwidth, 27 antennas, 0.1 h on source.}, with an RMS noise
in 10 min of about $45~\mu$Jy.  This giant leap in the sensitivity of
SKA will open up a whole new paradigm of radio astronomical research..

   Since SKA will be the premier radio interferometer in the future,
its high resolution capabilities should be maximized for astrometric
use and milliarcsec radio source imaging for the faint sources that
SKA will detect---without significant impact to the lower resolution
astronomical goals.  For this optimization, we recommend that SKA
long-baseline components should:
\begin{itemize}
\item {} have 25\% of its collecting area in these long baselines to
reach faint sources and to insure `in-beam' calibration up to 22 GHz.
\item {} contain at least 60 separate elements relatively uniformly
spaced between 1000 km and 5000 km from the core SKA.
\item {} observe simultaneously with at least ten independent
subarrays, each with at least six elements, with the core SKA.
\end{itemize}
\noindent
The reasons for these recommendations are given in the remainder of
this section. 

\subsection{The VLB Component}

   Since the core of SKA will be contained in a small area (even 150
km is considered small for astrometric use), the VLB component of SKA
will consist of many small elements located thousands of kilometers
from the core SKA.  Significant astrometric precision and high
resolution imaging can be obtained if the maximum baseline length is
about 5000 km.  Longer baselines would be desirable, but the gain in
astrometric accuracy does not increase as fast with longer baselines
because of the unavoidable use of low elevation observations.  The VLB
configuration should be well distributed in length and orientation, and
the core SKA should be located at one extremity of the VLB array to
maximize the sensitivity on the longer baselines.

   The VLB elements will be correlated only with the central element
(core SKA) since correlation among the outlying elements themselves
will have insignificant sensitivity.  This limits the (u-v) coverage
since the number of usable baselines varies only with the number of
VLB elements, rather than the square of the number.  A reasonable goal
is to have (u-v) coverage better than that for the VLBA with good
instantaneous beam characteristics for short observations.  We believe
that 60 elements, well distributed in distance and orientation from
the central core, is the minimum number for VLB elements to adequately
fill the (u-v) plane in order to obtain high dynamic range images of
complex objects.  Such high dynamic range images are also important
for astrometric consideration since all AGN (the typical radio
calibrator) are variable and their apparent core locations must be
accurately monitored.  Sixty elements is also a minimum number when
splitting the SKA-VLB capabilities into many independently observing
subarrays, see \S 2.4.

    It is expected that the low resolution part of SKA will be used
the majority of time for the many exciting low-resolution astronomical
experiments, and, thus, will not observe with the VLB elements.
However, the VLB elements alone, with at least 60 elements, will be a
useful instrument for simultaneous observations of relatively bright
sources ($>10$~mJy) for astrometric or other purposes.  Hence, we
envision two modes of operation of the VLB elements: 1) with the
entire SKA for accurate astrometric measurements and high quality
imaging of weak sources, and 2) in a stand-alone mode with
capabilities better than existing VLBI arrays on stronger sources.

    Although 5000 km is smaller than existing VLBI arrays, these
baselines will provide SKA with sufficient resolution to make high
resolution images and determine microarecsec astrometric accuracy.
However, SKA will undoubtedly be used in a VLBI mode with other
existing large telescopes on the earth to increase the baseline
length.  With several orbiting radio telescopes possibly operating
around 2020, the core SKA will be the crucial earth-based element for
baselines as long as 100,000 km.

\subsection {SKA-VLB Sensitivity}

    The SKA-VLB sensitivity depends on the amount of collecting area
in the VLB elements associated with SKA.  If $\eta$ is the fraction of
the SKA area which is contained in VLB elements more than 1000 km from
the core SKA, then the VLB (mas-scale size) sensitivity, $v$, with
respect to the nominal SKA sensitivity is $v\approx 0.8
\sqrt{\eta(1-\eta)}$.  (The estimated factor of 0.8 will depend on how
the core SKA is phased to obtain one large effective telescope.)  With
the present nominal SKA long baseline specification that 25\% of the
baselines should be longer than 150 km \cite{jon04}, we guess that
approximately 15\% of the baselines will be longer than 1000 km.  This
leads to a SKA-VLB sensitivity which is 28\% that of the SKA, giving
an RMS noise in 10 min of $0.35~\mu$Jy, and $0.05~\mu$Jy in 12 hours.
The stand-alone VLB elements alone have an RMS which is 15\% of the
entire SKA, or an RMS of $0.70~\mu$Jy in ten minutes.  With the
existing ground instruments the best sensitivity that is available for
VLBI (the EVN + phased-VLA + phased-Westerbork + one DSN-70m
telescope, with two polarizations, each at 1 GHz bandwidth, one bit
correlation) gives an RMS in 10 min is $25~\mu$Jy, a factor 70 less
sensitive than the SKA-VLB.

\subsection {SKA-VLB Calibration, Image Quality, Position Accuracy}

    The excellent SKA-VLB sensitivity can only be fully utilized by
accurate calibration of the data.  For baselines over 1000 km, the
dominant phase errors are associated with the large and rapidly moving
tropospheric and ionospheric phase screen over each telescope.  These
changes are smaller at kilometer baselines, although still a
problem.  Direct methods to measure and remove the effect of the
variable phase screen (water vapor radiometry, global satellite
transmission properties) have had limited success.  For the stronger
sources which can be detected within one coherence period (typically
five minutes at 8 GHz), self-calibration techniques can be used to
calibrate and image the source.  The minimum correlated peak flux
density that can be self-calibrated will depend on details about the
SKA-VLB array, but a reasonable minimum value is $5~\mu$Jy at 8 GHz.
The image quality can be quite good, and is often limited by
signal-to-noise considerations.  {\it However, positional information is
lost with self-calibration}.

    To determine the relative position of a radio source, or to image
a target source which cannot be detected within a coherence time, an
adhoc calibration scheme known as {\it phase referencing} is used.  A
relatively strong, small-diameter source (calibrator) which is within
a few degrees of the target, is also observed simultaneously with the
target if possible \cite{hon03}, or by quickly alternating
observations with the target.  The phase changes in the direction to
the calibrator for each telescope can then be determined, using the
self-calibration technique, and then applied to the target source as a
good approximation of the propagation errors in the direction to the
target.  This approximation decreases in accuracy, roughly, with the
calibrator-target separation.  For a one degree calibrator-target
separation for VLB observations, the residual phase errors in the
target limit the dynamic range of the image at 8 GHz is 30:1, or to a
position accuracy with respect to the calibrator or about 1/60 of the
resolution, which will be about 3 mas.

     However, with the superb sensitivity of the SKA-VLB and its
relatively large field of view (the area of sky for which images can
be instantaneously made), suitable calibrator sources will be
detectable within the field of view chosen for any target at all but
the highest planned SKA frequencies.  This `in-beam' calibration
technique will greatly enhance the SKA-VLB imaging quality and
astrometric accuracy for several reasons: The calibrators and targets
can be observed simultaneously; the calibrator target separations are
small; and with the use of several calibrators, the angular properties
of the phase screen over each telescope can be determined, to produce
a near perfect calibration in the direction of the target.

    An estimation of the in-beam calibration potential for the SKA-VLB
is as follows, using a frequency of 15 GHz.  (At the lower
frequencies, many tens of sources will be detectable).  We will take
the present SKA specification that the field of view is one degree
diameter at 1.4 GHz, or 0.1 degree at 15 GHz..  The density of sources
at 15 GHz with a correlated flux density greater than 5 mJy at a
baseline of 3000 km is about 1 source per square
degree\footnote{Source count at 8 GHz \cite{win93}, modified to 15
GHz, and assuming 30\% of flux density is in a core component.}.  A
reasonable extrapolation to lower flux densities is that the density
of sources increases inversely with flux density.  Assuming a
detection limit of $10~\mu$Jy at 15 GHz (using $\eta=0.15$), about 5
sources within the field of view can be detected.  At 22 GHz, there
are about 2 sources detectable above the expected detection level.
However, any decrease in the nominal sensitivity of SKA \cite{jon04}
will endanger calibration of the SKA-VLB at 22 GHz and higher
frequencies.  In fact, the assumption of $A/T=20000$ at 22 GHz maybe
optimistic by a factor of three.

    The ultimate dynamic range can be estimated as follows: Using two
calibrators, the closest about 1 degree from the target, a dynamic
range of 100:1 has been obtained \cite{fom04a} at 8 GHz, as shown in
Fig.~1.  With several calibrators available within a separation of a
few arcmin, a dynamic range, the changing phase gradient in
the vicinity of the calibrators and target can be determined and
removed.  This will lead to images after calibration with a dynamic
range of 1000:1 or better.  At 8 GHz with the SKA-VLB system, the
brightest source within the field of view is about 1 mJy; hence, a
dynamic range of 1000:1 would produce an effective noise of $1~\mu$Jy
near this source and about $0.1~\mu$Jy elsewhere in the field of view.
Although these dynamic range limits are still somewhat larger than
that expected receiver noise noise, extremely accurate images of weak
sources can be obtained.

     The use of close, multiple calibrators to targets not only
produces high quality images, but accurate positions with respect to
the calibrators.  In the example from Fig.~1, the positional accuracy
of the bright core of the radio source decreased from $35~\mu$sec to
$10~\mu$sec using two calibrators.  As a general rule, the astrometric
accuracy is equal to the resolution divided by the dynamic range of
the image.  Thus, the SKA-VLBA at 8 GHz, with a resolution of 3 mas and a
dynamic range of 1000:1 can determine positions to an accuracy of
about $3~\mu$as.  At this level, the precise positions of the
calibrators and their evolution of time must be also determined, and
will be discussed in \S4.x.

\medskip\noindent

\includegraphics[width=16pc]{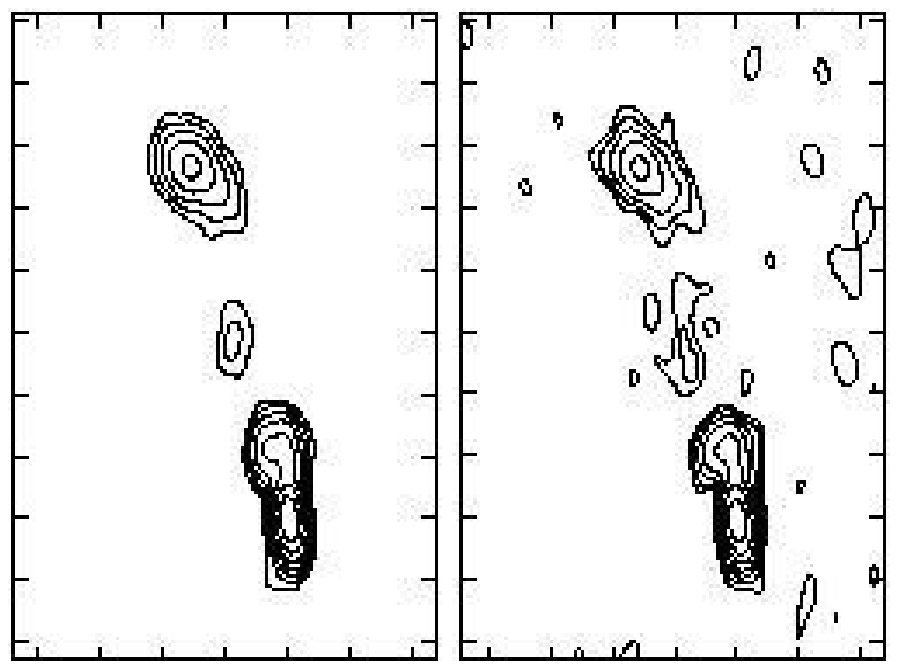}

\smallskip
\footnotesize
{\bf Fig 1:} (left) A radio source image after calibration using two
nearby calibrators, one $0.8^\circ$ to the east and another
$3.2^\circ$ to the west. (right) The radio source image after
calibration using only the nearer calibrator.  The minimum contour
level on both plots is 3 mJy and increase by a factor of 2.  The tick
marks are separated by 2 mas.

\medskip
\normalsize

    In conclusion, it is clear that if an appreciable component of SKA
is contained in elements which are more than 1000 km from the core
SKA, then milliarcsec-resolution images and microarcsec-accurate
positions can be obtained as long as in-beam calibrations are
possible.  With the present SKA design with $\eta=0.15$, it is
possible that such calibrations will be difficult at frequencies at or
above 20 GHz; hence we recommend that 25\% of the SKA collecting area
be located at least 1000 km from the core SKA.

\subsection {Global Astrometry}

    Absolute astrometric and geodetic experiments require observations
of relatively strong, small-diameter sources over the entire sky in as
short a time as possible in order to separate many astrometric and
geodetic effects by virtue of their different angular and temporal
behaviors.  Because of the small number of elements in current VLBI
arrays, such observations are only possible by serially observing
sources over the sky with the entire array, and this limits the
determination of simple tropospheric solutions to every 30 minutes to
one hour.  The resultant imprecise tropospheric solutions are the main
limitation to all-sky astrometric accuracy and derived geophysical
parameters.

    The present SKA specification includes the use of at least 10
subarrays.  This means that 10 independent, simultaneous VLB-type
observations, can be made on sources over a large part of the sky.
Because relatively strong and simple sources can be used for these
astrometric observations, short observations with modest (u-v)
coverage are sufficient.  Within about 5 minutes of observing time
(each subarray need only integrate on a strong source for about one
minute), the temporal and angular properties of the
tropospheric/ionospheric refraction screen over the SKA could be
accurately determined, and its effect removed.

    Hence, it is not primarily the sensitivity of the SKA which is
crucial to these absolute astrometric and geodetic observation, but
its ability to observe at least ten independent parts of the sky
simultaneously.  If we assume that the SKA-VLB component contains 60
elements, as needed for the imaging of complex sources, then each of
the ten independent SKA-VLB arrays will have only six baselines each.
However, this (u-v) coverage, although marginal, should be sufficient
when using strong, small-diameter sources, of which many will be
known.

    We estimate that the SKA-VLB, when used with at least ten
subarrays, would improve the absolute astrometric accuracy by at least
a factor of 10 over current day observations.  These all-sky
astrometric programs are also those which determine the earth
orientation and rotation, nutation, and dynamically link the quasar
reference frame to the planetary reference frame with observations of
spacecraft, beacons and solar-system objects and pulsars.

\section {RELATIVE ASTROMETRY: THE MOTION OF CELESTIAL OBJECTS:}

    By relative astrometric observations, we mean the determination of
the position and motion of a celestial object {\it with respect to
nearby background quasars}.  This type of astrometry will form the
bulk of the anticipated use with the SKA, and it is here that the {\it
in-beam} calibration method will be crucial in obtaining high
astrometric precision, as discussed in \S2.3 For a 1 mJy source,
observed over one hour with the SKA-VLB with our recommended array
properties, the potential astrometric accuracy is $3~\mu$as.  This
accuracy is sufficient to determine the parallax of any detected
galactic object and the proper motion and rotational speed of galaxies
within about 30 Mpc.

\subsection {Galactic Objects}

    The distance determination using trigonometric parallax to all
radio stars in our galaxy with a flux density greater than 0.5 mJy can
be determined with four six-hour SKA-VLB observations, spaced every
six months over a two-year period.  Distances to objects as weak as
$25~\mu$Jy can be measured within 1 kpc, while distances to stronger
sources can be done in much shorter integration times.  Several maser
transition lines are also available: OH at 1.6, 5, 6 GHz; CH3OH at
6.7, 12.2 and 23.2 GHz, H$_2$O at 22.2 GHz and, possibly SiO at 43
GHz.  Observations lower than 5 GHz are affected by the ionospheric
refraction, however, the wide SKA bandwidth will permit the removal of
the ionospheric component because of its $\nu^{-1}$ phase behavior
\cite{bri00}.  Thus, the distance of thousands of pulsars can be
obtained and comparison with dispersion measure and the rotation
measure of nearby calibrator sources are a tracer of the density,
temperature and turbulence of the interstellar medium.  The comparison
of the interferometric-derived positions and timing-derived positions
improve the modeling of pulsar properties, as well as aligning the
quasar inertial and the solar-system dynamic frames.

    Two experiments which determine the zero-point of the universal
distance scale can be significantly improved: measuring the distance
of stars in the the Hyades and Pleiades clusters which are faint and
need SKA sensitivity, and an improved calibration of the Cepheid
period/luminosity relationship.

    The proper motion of galactic objects are easily measured within a
period of a few years.  An example of the astrophysical ramifications
of such a measurement is illustrated in Fig.~2 which shows the proper
motion of Sgr A* over a 7 year period \cite{rei04}.  The residual
motion of Sgr A* of $<1$ km s$^{-1}$, from that expected from the
reflexive motion of the sun around the galactic center, gives a
minimum mass of the Sgr A* of $\sim10^6$ M$_\odot$, essentially requiring
a black hole.  With the SKA-VLB, the parallax could easily be
measured, providing a direct determination of the fundamental parameter,
$R_0$ to 1\% accuracy.  
Changing $R_0$ directly affects measured masses and luminosities of
almost all objects in the Milky Way, as well as for the Milky Way itself.

\medskip\noindent

\includegraphics[width=16pc]{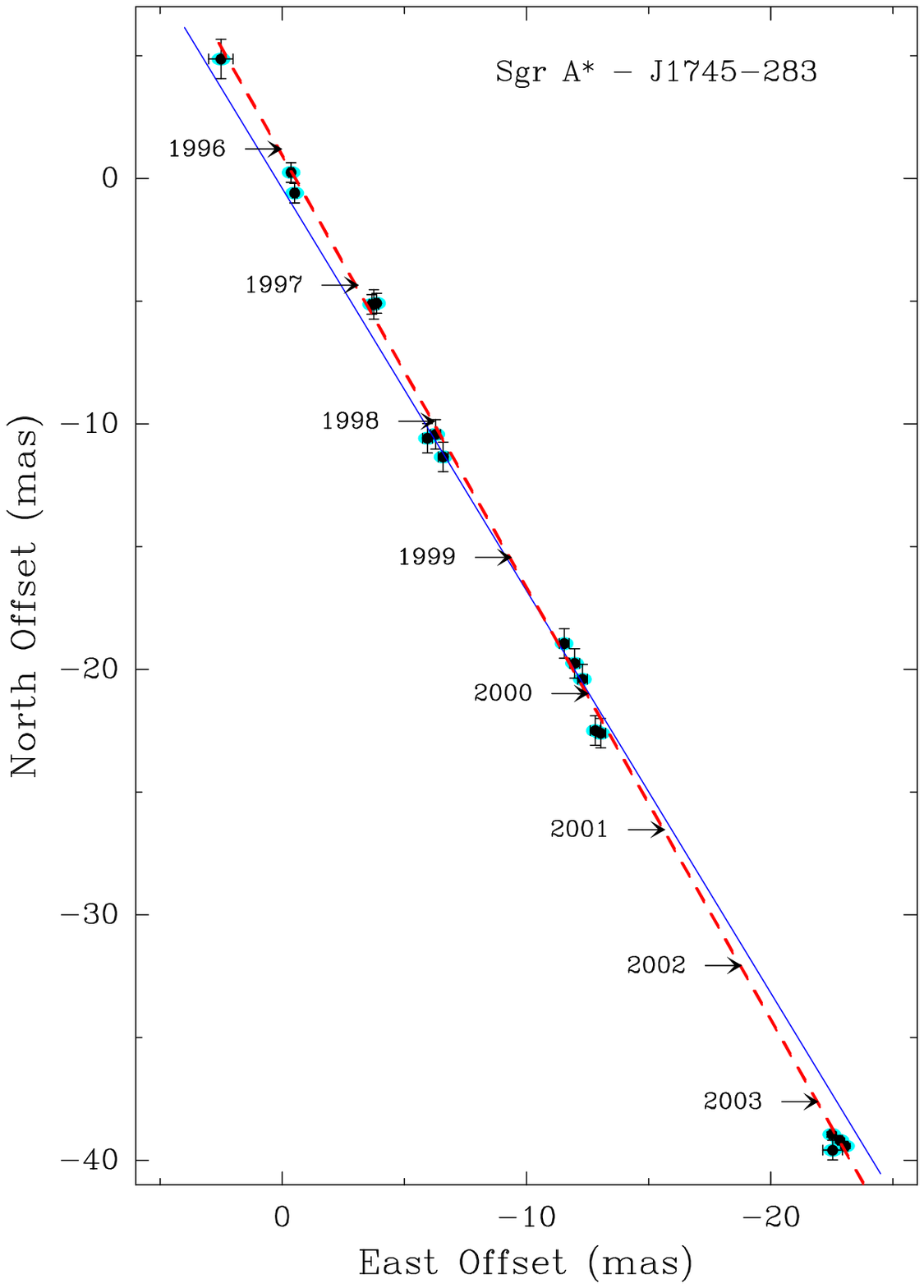}

\smallskip
\footnotesize
{\bf Fig 2:} The motion of Sgr A*, measured with respect to the quasar
J1745-283: The plotted points show the measured position and error
estimate of the location of Sgr A* between 1996.3 and 2003.5.  The
dashed line shows the best fit to the proper motion, and the solid
line shows the orientation of the galactic plane, arbitrarily aligned
with Sgr A* in 1999.6.  The apparent motion of Sgr A* is produced by
the solar motion around the galactic center, including the motion
perpendicular to the galactic plane plus a residual motion which
determines a lower mass limit of Sgr A*.  

\normalsize

\medskip 

    Other astrometric experiments with galactic objects are the stability
and dynamics of clusters, the use of maser tracers to determine the
turbulence and shocks of expelled material around dying stars or
infalling material around new stars.
\medskip\noindent

\includegraphics[width=16pc]{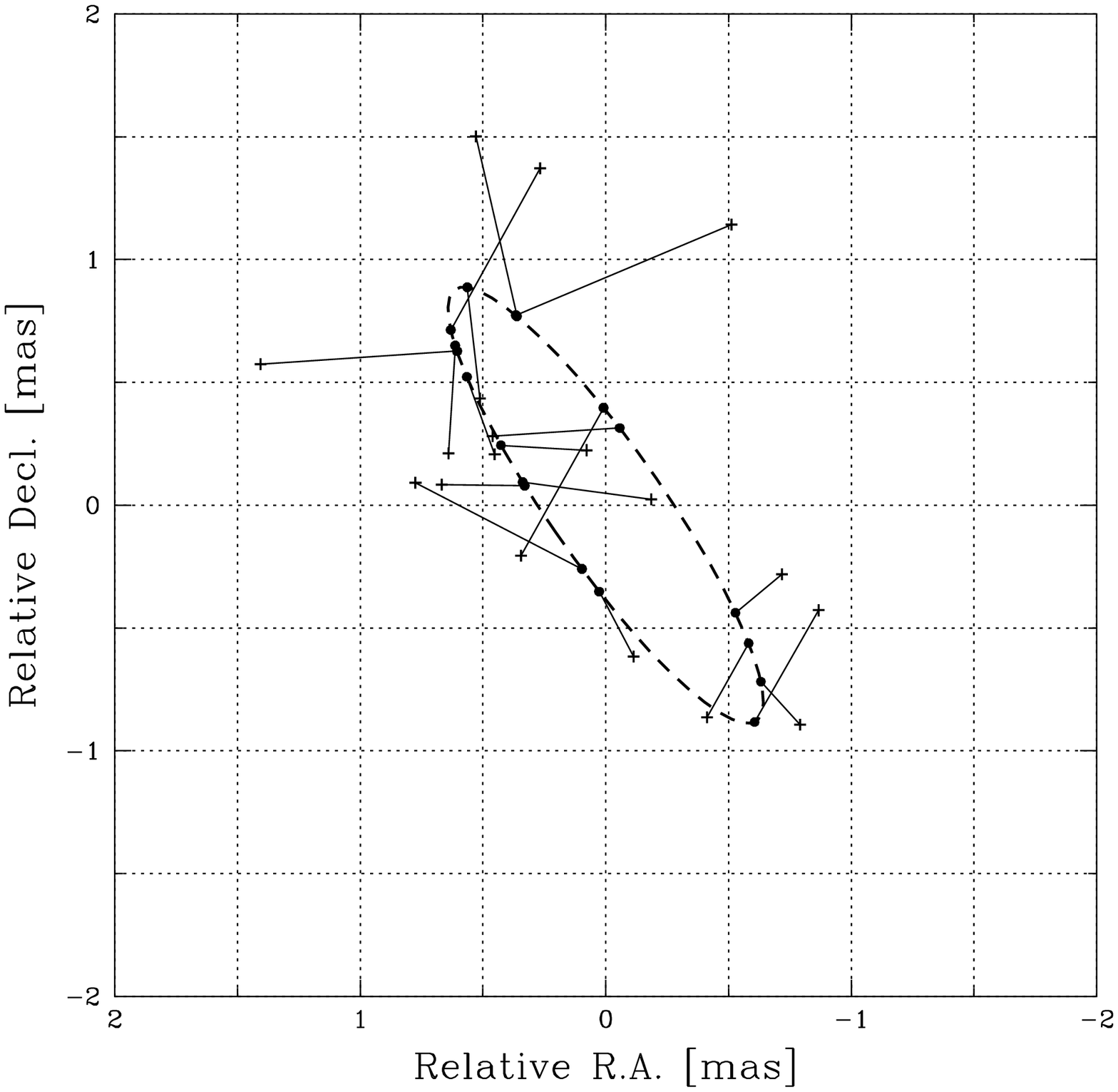}

\smallskip
\footnotesize
{\bf Fig 3:} The Orbital Motion of HR8703: The dashed line shows the
best fit orbital motion associated with HR8703 after removal of the
parallax and proper motion based on 4 years of observation.  The
plotted crosses show the measured residual positions, and the dots
show their expected location on the orbit with a 2.5-day period.

\medskip   
\normalsize

   The orbital motion in many binary systems will be readily
detectable; hence the mass function of many such systems can be
accurately determined.  An example of current day accuracy is given in
Fig.~3 for the RS~CVn binary, HR8703 (IM peg) \cite{ran04}.  This is
the reference star that the Gravity-Probe B mission is using to
measure the two gravitational precession terms predicted by General
Relativity as the spacecraft orbits the earth.  The figure shows the
residual orbital motion of the radio emission (associated with the
brighter star) after removal of the proper motion and parallax.  The
16 observations covered a period of 4 years and the orbital period is
2.5 days.  With the SKA-VLB, the orbit determination will be improved
by over a factor of 10 because of the much improved calibration and
sensitivity.  A similar experiment, the detection of Jupiter-sized
planets around stars up to a distance of 1 kpc, can be detected from
the planetary reflex motion of the star.

   X-ray binaries have properties which are similar to quasars and
other AGNs, but which evolve at a rate which is $10^7$ times faster
than AGNs.  Several examples are GRS1915+105 \cite{fen99}, SS433
\cite{par02,mio04} and Sco X-1 \cite{fom01a,fom01b}.  The motion of
the lobes and material in the jet and comparison with the x-ray
emission provide insights into the interaction of accretion disks
around black holes and neutron stars and the stability and lifetime of
jets.  All of these phenomena are also seen in AGN, but are more
accessible using observations of these galactic objects.  The imaging
and astrometric capabilities of the SKA-VLB will open up observations
to hundreds of these galactic micro-quasars.  Their rapid evolution
over a few minutes of time are barely accessible with current arrays,
but will be easily followed with SKA.

\subsection {Nearby Extragalactic Objects}

    Since most standard candles used for extragalactic distances are
tied to the distance of the LMC, direct measurement of this galaxy and
others out to the Virgo Cluster could improve the distance scale
calibrations to 1\% accuracy.  Unfortunately, direct parallax
measurements are only just reachable for the LMC by the SKA.  However,
the proper motion associated with the rotation of stars in galaxies as
far as the Virgo Cluster are about 5 to $100~\mu$as/yr, and can be
accurately measured by SKA over a few years.  These angular velocities
can be used as a direct measurement of the galaxy distance by
comparing them with the known rotational velocities obtained by a
variety of methods.  Fig.~4 illustrate the results from recent VLBA
observations \cite{bru03} which have measured the proper motion
between two H$_2$O maser sources that lie on the opposite sides of
M33.  The derived distance estimate is already better than 15\%.
Similar observations of masers by SKA will detect many more examples
and determine the distance scale to 1\% accuracy out to 30 Mpc.

   The luminous OH and H$_2$O megamasers are associated with the dense
material in accretion disks of AGN, and the study of their motion can
reveal the dynamics of the accretion disk and the nuclear (black hole)
mass.  The high velocities expected in the latter stages of
galaxy-mergers, as in Arp220, may be measurable out to 100 Mpc with
SKA.

\medskip\noindent

\includegraphics[width=16pc]{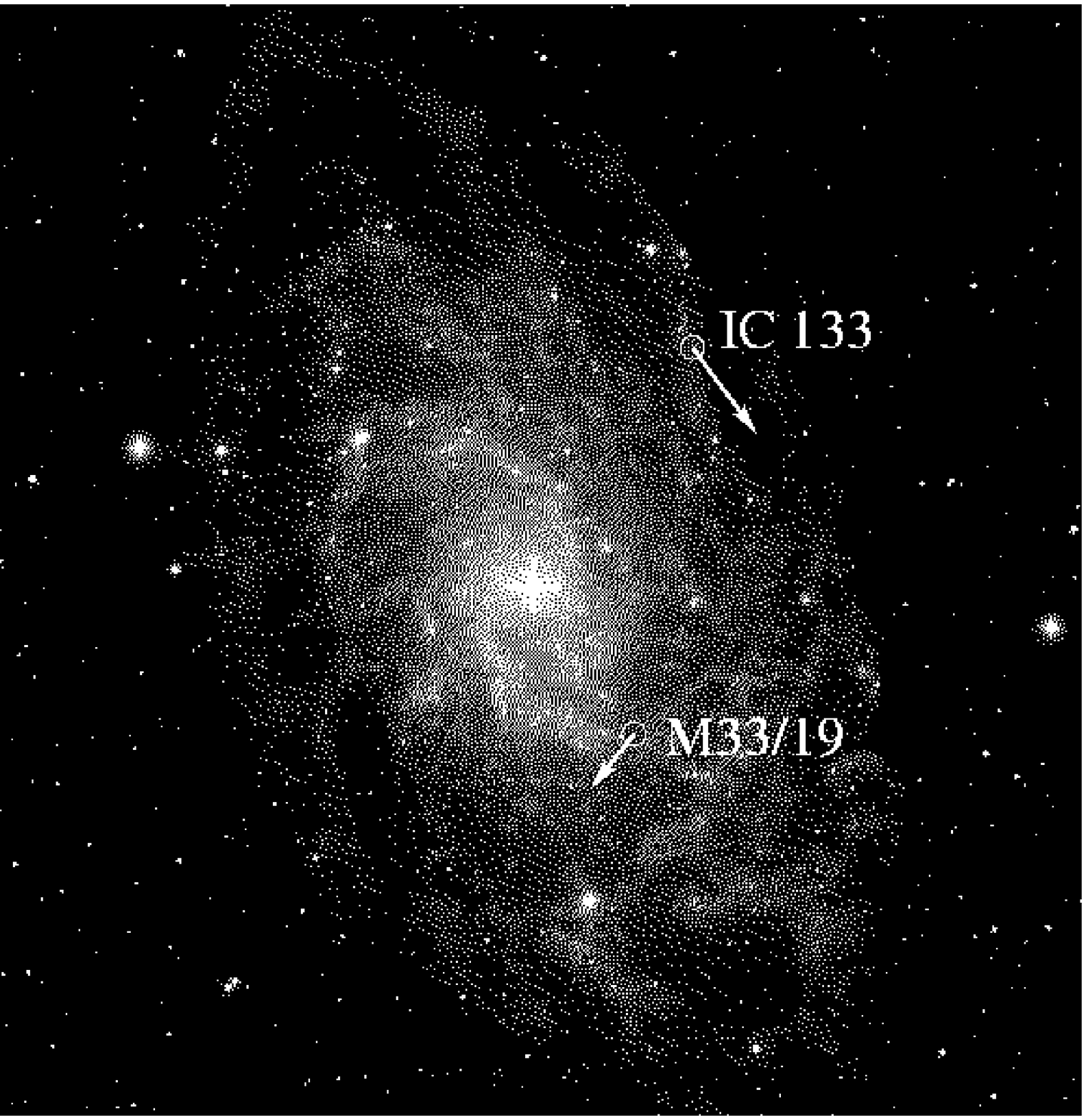}

\smallskip
\footnotesize
{\bf Fig 4:} The Motion of Two Maser Sources in M33: The arrows
show the expected motion of the maser source IC133 (100 km/s) and
M33/19 (62 km/s) with respect to the radio core, derived from the
known rotation model for the galaxy.  The radio observations of the
relative motion of the two maser sources were made over a 2.5 year
period and give a distance uncertainty to M33 of 15\%.

\medskip
\normalsize
\subsection {Distant Extragalactic Objects}

   The milliarcsec imaging of quasars and AGN, although an important
aspect of the SKA, is not the main concern in the chapter.  Detailed
images of the fainter radio sources which are dominated by the star
formation process, will doubtlessly uncover the wealth of complexity
in these distant galaxies.  Comparison with observation at sub-mm,
infrared, optical, ultraviolet, X-ray and gamma ray energies will be
an important aspect in the understanding of galaxy formation and
evolution at early epochs.  Recent VLA and MERLIN observations are
beginning to show the complexity of the radio and optical emission
from these distant galaxies \cite{fom04b,mux04}

   However, one feature of quasars and AGNs as astrometric calibrators
is important: they are variable.  The radio emission from the smallest
component of a radio source, called the radio core, come from the base
of the jet which is formed by an accretion disk around a massive black
hole.  The jet is optically thick at SKA radio frequencies and its
apparent position is a function of frequency.  The massive outflows
and shocks in the jet further change the intensity and the structure
of the radio emission.  Hence, the position of the so-called radio
core is variable by about 0.05 mas in most AGN \cite{sud03}.  Thus,
the calibrators used to determine the SKA astrometric precision to
better than $10~\mu$sec as subject to the calibrator position jitter
which is somewhat larger.  In order to reach the intended angular
precision, the change in position of the calibrators must be
determined.  With knowledge of the evolution of radio cores and the
monitoring of a basic set of primary calibrators, we believe that the
calibrator grid can be determined to $10~\mu$as or better.  The use of
many calibrators in the field of view also diminished the net effect
of position jitter.

\section {ABSOLUTE ASTROMETRY AND GEODESY USING THE SKA}

   By absolute astrometry we mean the determination of the true
positions of a radio source defined in a well-defined inertial-frame
system.  The SKA by virtue of its ability to observe sources over the
sky in a very short period of time, will be as competitive in this
endeavor as any of the currently planned orbiting optical astrometric
satellites.

\subsection {The Fundamental Reference Frame}

   The fundamental reference frame is currently defined by the quasars
which are so distant that they are essentially fixed in the sky.
Based on millions of radio observations of quasars around the sky from
the 1970's, an inertial reference frame tied to the quasars was
established in 1995, the International Celestial Reference Frame
(ICRF) \cite{ma98}.  The frame origin is tied to the solar system
barycenter and its orientation in space is currently known to
$50~\mu$as accuracy.  The main limitation of the accuracy is the
tropospheric refraction which affects radio observations.  The
large-scale tropospheric refraction can be estimated by observing many
radio sources over the sky in a short period of time.  However, at
present the determination of the global troposphere properties can
only be estimated in about one hour, and smaller angular-scale
variations can not determined.

   The SKA, by using observations in ten subarrays, on strong radio
sources around the sky, will determine the tropospheric properties on
time-scales which may be as short at five minutes.  The resultant
tropospheric models will then include not only the changes of the
global property of the refraction, but deviations in its azimuth and
elevation dependence.  Thus, the stability of the reference frame
should be reduced by about a factor of 10 from current day precision,
so that the non-stationarity of the quasars themselves, may become the
dominant error.

\subsection {Geodetic, Earth and Solar System Dynamics}

   In reaching toward an inertial frame which is consistent to
$<10~\mu$as, all aspects of the motion of the SKA on the earth's crust
as it spins and wobbles in space in a solar orbit which is perturbed
by the planets and other solar system bodies, while moving around the
center of the galaxy, must be determined.  This information is used by
geodesists to monitor the motion and stresses within the earth, and
the interaction of its spin and nutation with terrestrial phenomena.

   The dynamics of the solar system bodies, through their effect on
the earth and the SKA, can be tied precisely to the quasar inertial
frame.  SKA observations of spacecraft and beacons in space will aid
in this tie.  Comparison of the interferometric position of pulsars
(tied to the quasar frame) and their timing-based position (tied to
the motion of the earth) are also important.  Short-term fluctuations
in the earth rotation will be easily determined from SKA, and these
changes can be associated with deformations (earthquakes?) in
the earth.

   When the core SKA is engaged in low resolution astronomical
observations, the VLB elements alone (or possibly with a small portion
of the core SKA) could be used for many of these all-sky astrometric
observations which use only the stronger sources.  This is another
reason to have a sufficiently large VLB component of SKA so that it
can stand alone and provide observations of this type.

\subsection {Tests of Fundamental Physics}

   The general relativistic effects of the interaction of space and
time become important in the determination of an accurate inertial
system, and the future inertial frame must be fully four-dimensional
where space and time are intertwined in accordance with GR.  The
monitoring of spacecraft trajectories, the bending of radio waves from
the sun and Jupiter \cite{leb95,fom03}, and the delay of signals from
spacecraft can be used as precision tests of Einstein's formulation of
gravity in the low field region of the solar system.

   One of the most important projects of SKA, the detection of low
frequency gravity waves from changes in the global timing properties
on many pulsars, is dependent on the precise knowledge of the inertial
frame, tied to the quasars.  A slight deviation between the quasar
inertial frame and the dynamical frame associated with the solar
system will produce a global timing analysis residual which can mimic
that of a gravity wave passing through the solar system.

\section {SUMMARY}

   The SKA, with about 25\% of its collecting area in elements from
1000 km to 5000 km from the its core, will provide astrometric
capabilities which are a factor of 10 more accurate than now
available.  These high resolution observations will concentrate at the
higher SKA frequencies, from 1 GHz to 23 GHz, although even higher
frequencies would be useful.

   The sensitivity of SKA will permit the measurement of the motion of
a large number of relatively weak and interesting radio sources to
the microarcsec level, and provide excellent calibration of
tropospheric refraction for good image quality.  The versatility of
SKA to observe objects simultaneously over the sky will improve the
fundamental astrometric and geodetic experiments to an accuracy of
$10~\mu$Jy.

   We believe that the scientific payoff, as outlined in this paper
for high resolution capabilities of SKA, will more than compensate for
the slight loss of sensitivity for the many exciting, but lower
resolution, SKA experiments.  The data links between the core SKA and
the VLB elements will not be expensive or technologically challenging,
and the data rates associated with most SKA-VLB experiments will be
less than envisioned in many other SKA key science projects.

\end{document}